\documentclass[sigconf,natbib=false]{acmart}
\AtBeginDocument{%
  }
\usepackage{todonotes}

\setcopyright{acmlicensed}

\copyrightyear{2026}
\acmYear{2026}
\acmDOI{XXXXXXX.XXXXXXX}
\acmConference[SAC'26]{The 41st ACM/SIGAPP Symposium on Applied Computing}{March 23--27, 2026}{Thessaloniki, Greece}
\acmISBN{979-X-XXXX-XXXX-X/26/03}

\RequirePackage[
  datamodel=acmdatamodel,
  style=acmnumeric,
  ]{biblatex}

\begin{document}

\title{Framework Matters: Energy Efficiency of UI Automation Testing Frameworks}
\renewcommand{\shorttitle}{Framework Matters}


\author{Timmie M. R. Lagermann}
\email{timmiel@ruc.dk}
\orcid{0009-0002-4728-0305}
\affiliation{%
  \institution{Roskilde University}
  \city{Roskilde}
  \country{Denmark}
}

\author{Kristina Sophia Carter}
\email{kcarter@ruc.dk}
\orcid{}
\affiliation{%
  \institution{Roskilde University}
  \city{Roskilde}
  \country{Denmark}
}

\author{Su Mei Gwen Ho}
\email{smgho@ruc.dk}
\orcid{}
\affiliation{%
  \institution{Roskilde University}
  \city{Roskilde}
  \country{Denmark}
}

\author{Luís Cruz}
\email{L.Cruz@tudelft.nl}
\orcid{}
\affiliation{%
  \institution{Delft University of Technology}
  \city{Delft}
  \country{Netherlands}
}

\author{Kerstin Eder}
\email{Kerstin.Eder@bristol.ac.uk}
\orcid{}
\affiliation{%
  \institution{University of Bristol}
  \city{Bristol}
  \country{England}
}

\author{Maja H. Kirkeby}
\email{majaht@ruc.dk}
\orcid{}
\affiliation{%
  \institution{Roskilde University}
  \city{Roskilde}
  \country{Denmark}
}

\renewcommand{\shortauthors}{T. Lagermann et al.}


\begin{abstract}
We examine per action energy consumption across four web user interface (UI) automation testing frameworks to determine whether consistent tendencies can guide energy-aware test design. Using a controlled client–server setup with external power metering, we repeat each UI action (refresh, click variants, checkbox, drag\&drop, input-text, scroll) 35 times. Across each of the actions, energy costs vary by both framework and action. Puppeteer is the most efficient for left-click, right-click, double-click, checkbox, and input-text; Selenium is the most efficient for refresh and scroll; Nightwatch is generally the least energy efficient. 
The energy cost of performing the same action varied by up to a factor of six depending on the framework. This indicates that providing transparency of energy consumption for UI automation testing frameworks allows developers to make informed, energy-aware decisions when testing a specific UI action.
\end{abstract}



\maketitle

\section{Introduction}
Energy consumption has become a growing concern in software engineering, motivating research into sustainable and energy efficient development practices \cite{Zaidman2024,Verdecchia2025}. While considerable effort has been devoted to measuring energy use at the application and algorithmic levels \cite{Pereira2021,Gordillo2024}, less attention has been given to the energy implications of the tools and frameworks that underpin software development and testing \cite{Cruz2019}.

In recent years, a growing body of work has sought to make software energy behavior observable and actionable for developers. Academic approaches such as EDATA \cite{Blokland2025} and the energy-vampire debugging methodology \cite{Roque2025} illustrate this movement toward practical energy transparency, while industrial platforms such as GreenFrame \cite{GreenFrame2025} extend it by estimating the carbon footprint of web applications through browser automation. Within the Web domain, Macedo et al.~\cite{Macedo2020} demonstrated that browser choice (Chrome vs. Firefox) affects measured energy consumption when user interactions are automated with Selenium. Despite such progress, existing approaches remain coarse-grained: they attribute all measured energy to the application under test and overlook the influence of the automation framework itself.

From a research perspective, this gap aligns with what Verdecchia et al.~\cite{Verdecchia2025} define as energy-aware testing: a new frontier in software engineering where the testing processes, environments, and tools themselves become first-class subjects of energy analysis. UI automation frameworks occupy a central position in this space. They are widely adopted to simulate user interactions in web applications, valued for their repeatability, integration into continuous delivery pipelines, and cross-platform capabilities. 
. Yet, their own energy behavior—and the extent to which internal design choices such as synchronization or driver communication affect test execution—remains poorly understood.

Evidence from the mobile domain reinforces this concern. Cruz and Abreu~\cite{Cruz2019} demonstrated that the choice of UI testing framework can substantially influence measured energy consumption, and conclude that the
energy consumption of mobile UI automation frameworks should
be factored out to avoid affecting results of energy tests.
Extending this line of inquiry to computers, our study systematically examines the energy consumption of scripted UI actions across multiple testing frameworks for web software on computers to identify and isolate framework-induced effects that may bias sustainability assessments. This is particularly important in the web context because automated UI testing is widespread in modern development pipelines in web development. If the frameworks themselves introduce significant and inconsistent energy overhead, it can distort energy measurements and mislead optimization within web development. The goal is therefore to investigate the energy impact of using web UI automation testing frameworks.

We address the following research questions:
\begin{itemize}
    \item {\texttt{RQ1}}: Do different UI actions vary in their energy consumption within a single UI automation testing framework? \label{RQ:actions}
    \item {\texttt{RQ2}}: Does the choice of UI automation testing framework affect energy use for the same UI action? \label{RQ:frameworks}
    \item {\texttt{RQ3}}: Is there an interaction between UI automation testing framework and UI action type in terms of energy usage? \label{RQ:interaction}
\end{itemize}

The four frameworks investigated in this study are: Nightwatch\footnote{\url{https://nightwatchjs.org}}, Playwright\footnote{\url{https://playwright.dev}}, Puppeteer\footnote{\url{https://pptr.dev}} and Selenium\footnote{\url{https://www.selenium.dev}}. The selection of the frameworks will be presented in section \ref{subsec:inclusion} while the implication on the results from the choice of the frameworks will be discussed in section \ref{subsec:framework-dependent-energy-behavior}.
Our results show that UI actions differ significantly in their energy consumption, with drag\&drop and input-text emerging as the most energy-intensive actions across the four frameworks. 
In contrast, actions such as refresh, left-click and right-click showed relatively low and consistent energy profiles. The energy cost of performing the same action varied by up to a factor of six depending on the framework.
Among the tools tested, Puppeteer was the most energy efficient overall, followed by Selenium and Playwright, while Nightwatch consistently exhibited the highest energy usage. Notably, no single framework was uniformly optimal across all actions. 
Therefore, energy measurements obtained using the selected UI automation frameworks should be interpreted with care, as both framework behavior and implementation choices can introduce significant variation. Failing to account for or align these factors may lead to misleading comparisons, incorrect conclusions about energy efficiency, and irreproducible results across tools or test suites.
The main contributions of this paper are:
\begin{itemize}
\item{A demonstration that the energy cost of UI actions differs not only across frameworks, but also depending on the specific type of interaction performed.}
\item{Evidence that no single automation framework is consistently energy efficient across all types of UI actions.}
\item{Data to substantiate that different implementation strategies for the same UI behavior can lead to substantial differences in energy consumption.}
\end{itemize}

\section{Related work}

\subsection{Energy-Aware Software Testing}
Testing has been identified as a non-trivial contributor to software’s energy footprint. Zaidman~\cite{Zaidman2024} quantifies testing energy consumption in CI pipelines, and Misu et al.~\cite{Misu2025} show that test smells 
impact energy consumption has not been investigated yet.
correlate with higher energy use, making test quality itself an energy factor. Framing these findings, Verdecchia et al.~\cite{Verdecchia2025} formalize energy-aware testing and call for methods that consider the energy impact of testing processes and tooling. Our work answers this call by focusing on automation frameworks as a potential source of systematic variation in measured energy.

\subsection{Frameworks as Execution Environments}
The notion that the choice of execution environment influences energy efficiency has been well established at the programming language level.
Pereira et al.~\cite{Pereira2021} ranked 27 languages by energy efficiency, revealing that compiler and runtime differences can outweigh algorithmic optimizations.
Gordillo et al.~\cite{Gordillo2024} corroborated these results using hardware-based measurements, confirming that language design and runtime behavior have systematic effects on energy consumption.
Automation frameworks, much like languages, encapsulate execution logic and interact with system resources in distinct ways.
This analogy motivates our investigation into whether such framework-level differences also translate into measurable energy variation.

\subsection{Automation and Testing Frameworks}
Framework choice has been shown to influence both runtime characteristics and energy outcomes. In mobile settings, Cruz and Abreu~\cite{Cruz2019} report energy overhead differences exceeding 2000\% across UI automation frameworks, concluding that the framework can overshadow the system-under-test. 

In the web domain, Garcia~\cite{Garcia2024} describes the architectural differences of Selenium, Cypress and Playwright, but does not evaluate the frameworks. 
We extend this line by providing fine-grained, per-action energy comparisons across web automation frameworks under controlled conditions.

\section{Experiment Planning}

In this section, we will provide a general overview of the goal, the control used and the variables considered. These questions form the foundation for our experimental design and subsequent data analysis.



\subsection{Experimental design}
We employed a full-factorial within-subject design to compare the energy consumption of UI test actions across multiple frameworks. Each combination of framework and action was executed on the same physical hardware setup, eliminating hardware-induced variability.
No randomization or counterbalancing was used; actions were executed in a deterministic order defined by a fixed test script. This allowed for complete coverage of the framework–action matrix while simplifying replication. All the frameworks had coverage collection OFF to lower their energy usage. Potential ordering effects are addressed in the validity discussion in section \ref{subsec:internal-validity} and \ref{subsec:external-validity}.

\subsection{Inclusion and Exclusion criteria}
\label{subsec:inclusion}
This study evaluates four UI automation testing frameworks: Selenium, Nightwatch, Playwright, and Puppeteer. These frameworks are the subject of the study. Each framework was selected based on its popularity, community support, and compatibility with our technical requirements.
Inclusion criteria were as follows:
\begin{itemize}
\item{The framework must be widely used in practice, as indicated by GitHub stars, amount of subscribing GitHub Users, and Stack Overflow tag frequency. }
\item{It must support JavaScript-based test execution, since all test scripts in this study are written in JavaScript. JavaScript was chosen over other languages (e.g. Python) to minimize variance in language-level energy consumption, as suggested by Pereira et al.~\cite{Pereira2021}.}
\end{itemize}

Exclusion criteria were applied to ensure technical compatibility and data validity:
\begin{itemize}
    \item {The framework must be installable and executable on Raspberry Pi devices.}
    \item {It must be compatible with Chromium-based browsers, as all tests were executed in a headless Chromium environment. Chromium was selected for its low overhead and minimal memory usage.}
    \item {The framework must produce explicit and reliable failure logging. Frameworks that silently skipped steps or suppressed failure messages were excluded to ensure that measured energy consumption reflects actual test execution.}
\end{itemize}

These criteria collectively ensured that the selected frameworks were both relevant, directly comparable, and practically feasible.
The objects of this experiment are the individual UI actions automated by each framework. We define a UI action as a discrete user-initiated interaction, such as clicking a button or entering text. Actions were selected for their relevance to realistic user behavior (rather than efficient computation) and their ability to be implemented consistently across frameworks for direct comparability. Idle actions were excluded due to definitional ambiguity and implementation inconsistency. The overhead of starting up and closing down the frameworks were also measured, and the average energy usage of the overhead were subtracted from the energy cost of a given action for said framework. In this paper we consider overhead as the combined energy consumption of launching our script, which utilizes the different frameworks through JavaScript in order to open the web site, wait and close the framework again.
The selected actions represent fundamental building blocks of typical user interactions and were defined as follows. 
\begin{itemize}
    \item {refresh}
    \item {left-click}
    \item {right-click}
    \item {double-click}
    \item {checkbox}
    \item {drag\&drop}
    \item {input-text}
    \item {scroll}
\end{itemize}

However, drag\&drop is excluded for Puppeteer, which only supports drag\&drop by coordinate-based interaction and not by direct manipulation of Document Object Model (DOM) elements as the other frameworks do\footnote{\url{https://pptr.dev/api/puppeteer.mouse.draganddrop}}. Each of the actions were repeated differently to obtain enough measurements to do analysis upon, the amount of repetitions depended solely on the action and thus were the same across all the frameworks. In order to get enough measurements each of the action/frameworks experiments were furthermore repeated 35 times, with varying wait times for each framework, for the specific amount of repetitions and wait times see the tied repository\footnote{\url{https://github.com/Temmerln/Framework-Matters.git}}.

\subsection{Instrumentation}

The test environment was instrumented to measure energy consumption precisely and consistently across frameworks and UI actions.
The system under test was a Raspberry Pi 4B, running test cases one framework at a time. Each UI automation framework was installed on a dedicated clean SD card image, ensuring full isolation and avoiding cross-configuration interference. All tests were executed in headless Chromium, and browser caches were cleared between test cycles to eliminate carry-over effects.
Power was supplied and measured using a Siglent SPD3303X-E programmable laboratory grade power supply configured to deliver a constant 5V with a varying current (maximum~3.2A). Its voltage accuracy is ±0.05\% of the reading plus two digits, and the current accuracy is ±0.5\% of the reading plus two digits. The Siglent device logged power, voltage, current, and timestamped samples.
Test execution was coordinated from a control laptop connected via ethernet which launched the appropriate test scripts and collected Siglents measurement collection using Test Controller. While the Siglent logged the power trace, the Raspberry Pi itself logged the test duration and average core CPU temperature, recorded immediately before and after each test cycle. This dual logging setup allowed us to detect thermal drift, identify start and stop times of test executions, and assess whether background processes might have affected energy readings.
To eliminate external variability, a second Raspberry Pi acted as a minimal local server, hosting a static test website. The entire setup was connected through a network switch with four active ports, one for each of the connected entities. The switch, the control laptop, and the server were independently powered, and only the client Raspberry Pi drew power from the Siglent unit. The specific technical specifications for the instrumentation and the software can be found in the online repository\footnote{\url{https://github.com/Temmerln/Framework-Matters.git}}.

\subsection{Analysis Procedure}

This subsection outlines the methodology used to process energy measurements and perform statistical analysis to answer the research questions defined earlier.

\subsubsection{Data Processing}
Energy consumption was extracted using an eight step method combining power thresholding, log alignment, and trapezoidal integration. The procedure filtered warm-up noise, identified active execution periods using a calibrated power threshold, removed short-duration and timing-inconsistent spikes, and aligned intervals to framework logs. Each measurement was then annotated with an action ID and type. A similar approach was used to identify the average overhead for launching each of the UI frameworks, which are thus subtracted to identify the energy consumption.

\begin{figure}[h]
  \centering
  \includegraphics[width=\linewidth]{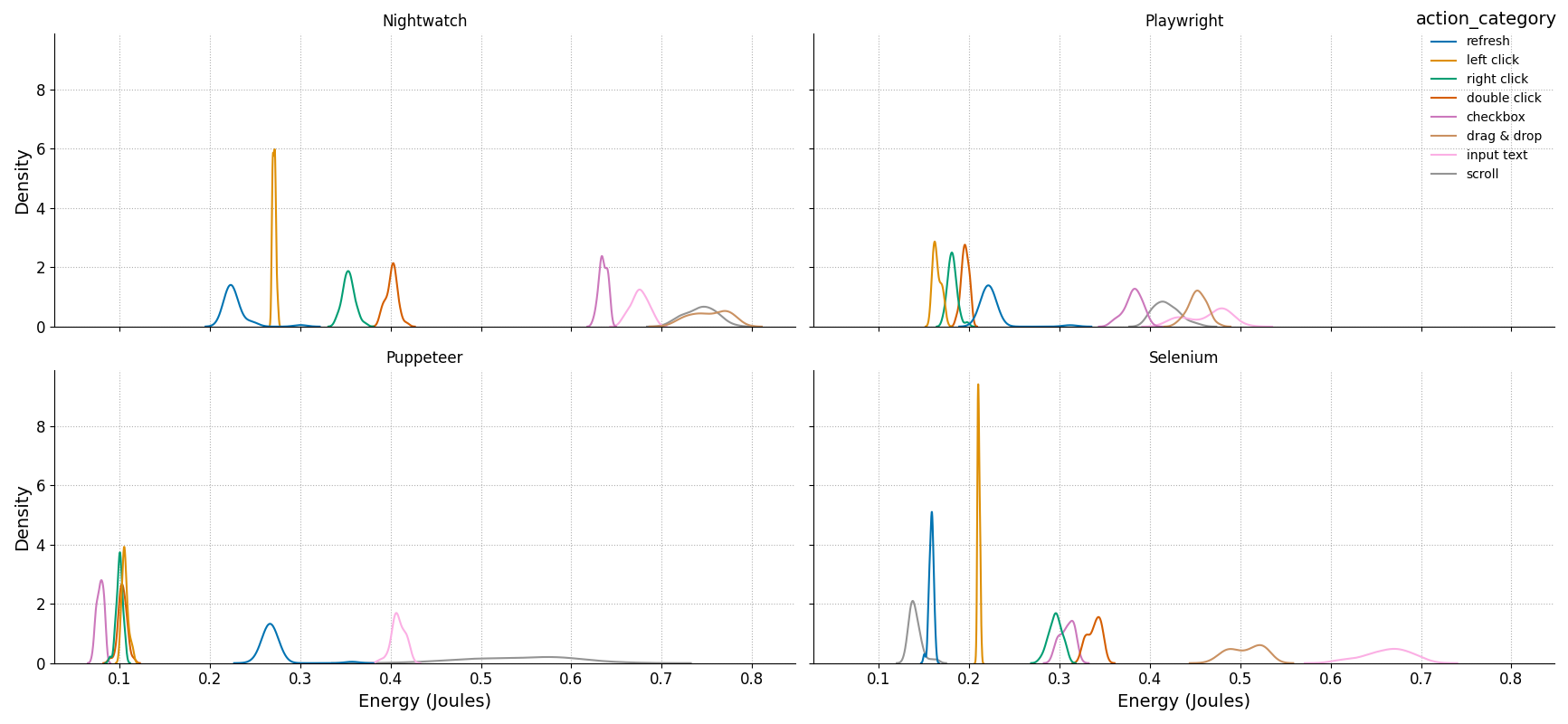}
  \caption{Visual verification of distribution shapes for energy consumption across framework action combinations. The plots reveal that many of the framework action combinations deviate from normality.}
  \label{fig:action_distributions}
  \Description{}
\end{figure}

\subsubsection{Statistical Analysis}
Statistical tests were applied based on the assumptions satisfied by the data. The analysis for RQ1, RQ2, and RQ3 focuses on the eight different UI actions.
The assumption of normality was tested for each of the framework action combinations using visual verification.  Figure~\ref{fig:action_distributions} shows that many actions, regardless UI automation testing framework, are not normally distributed. The visual inspection was supplemented by two complementary normality tests:
The Shapiro-Wilk test, suited for detecting deviations from normality in small to medium sized samples, and the Anderson-Darling test, sensitive to tail discrepancies in distributions.
Out of 31 testable framework action combinations (drag\&drop for puppeteer were not eligible test):

\begin{itemize}
    \item{20 combinations (64.5\%) passed the two complementary normality tests,}
    \item{1 combination (3.2\%) passed only one test,}
    \item{10 combinations (32.3\%) failed both tests.}
\end{itemize}

\begin{figure}[h]
  \centering
  \includegraphics[width=\linewidth]{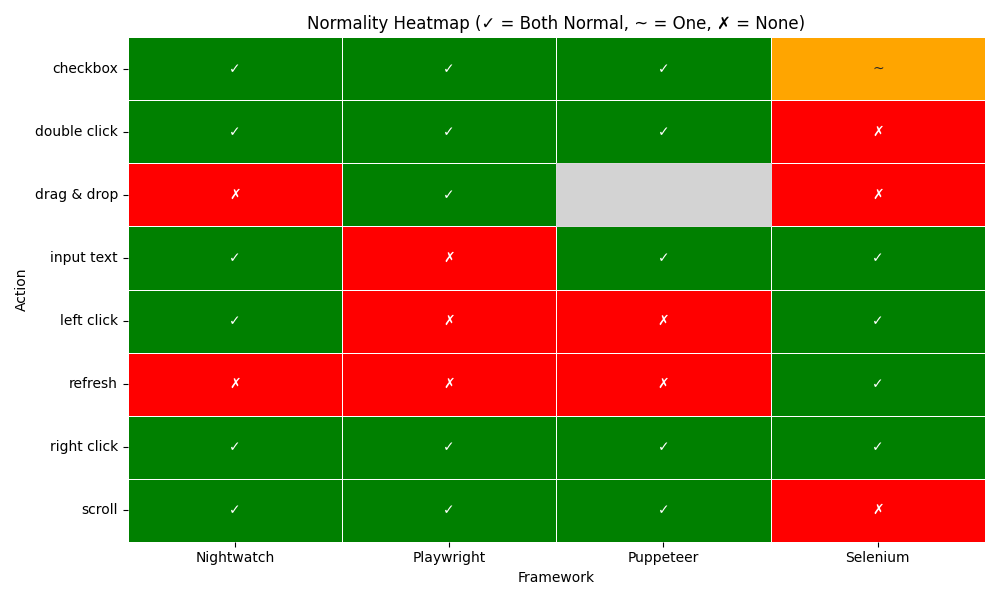}
  \caption{Visual summary of Shapiro-Wilk and Anderson-Darling tests. Green indicates a pass in both cases. Red indicates a failure in both cases. Yellow indicates a pass for one test and fail in the other. Gray indicates a non-posible action.}
  \label{fig:normaility_heatmap}
  \Description{}
\end{figure}

While a majority exhibited normal-like characteristics, a substantial portion failed one or both tests, indicating inconsistent outcomes. To avoid violating statistical assumptions the solution is to apply non-parametric methods for all the subsequent comparisons. The results of the Shapiro-Wilk test and the Anderson-Darling test are summarised in Figure \ref{fig:normaility_heatmap}. Here, it is worth noting that only the right-click action passed both normality tests across all the UI Automation Testing frameworks. Given these inconsistencies, we opted to use non-parametric statistical methods for all subsequent hypothesis testing to avoid violating assumptions of normality. The following non-parametric tests were adopted for all the subsequent hypothesis tests: 
For RQ1 (framework comparison for the same action) and RQ2 (comparison of actions framework by framework): We used the Kruskal-Wallis tests with Dunn’s post hoc test for pairwise comparisons when significant effects were found. RQ3 (interaction between framework and action): To test whether the effect of the framework varies by action, we applied Aligned Rank Transform (ART) ANOVA in R to test the interaction effects for non-normal conditions.

Our energy calculations are summarized in Equations~\ref{eq:action_total},~\ref{eq:overhead_total} and~\ref{eq:sleep}. $E_{action\_total}$ is the total energy of the action/framework combination (with the action specific repetitions). $E_{action}$ is the energy of the action, without the energy usage of the overhead/and passive energy consumption of having the machine running. $E_{overhead\_total}$ is the combined energy of the framework overhead and the passive energy consumption of having the machine running. $E_{sleep}$ is the sleeping power usage $P_{sleep}$ times the time t, it takes for the action to finish. While $E_{overhead}$ is the energy it takes to open and close the framework.
The goal for our experiments is to identify the, $P_{sleep}$, $E_{overhead}$ and $E_{action}$.

\begin{equation}
    E_{action\_total} = E_{action} + E_{overhead\_total}
    \label{eq:action_total}
\end{equation}

\begin{equation}
    E_{overhead\_total} = E_{sleep} + E_{overhead} 
    \label{eq:overhead_total}
\end{equation}

\begin{equation}
    E_{sleep} = P_{sleep} + t
    \label{eq:sleep}
\end{equation}

\section{Results}\label{results}

The following presents the results for energy of the respective overhead of the different frameworks and the findings in response to RQ1, RQ2 and RQ3.

\subsection{Overhead}
In order to calculate the energy usage of each UI action, we observed the baseline power consumption of the device under inspection during active sleep is 1.85~W; the same for all the frameworks. The overhead or mean energy consumption for opening and closing the respective frameworks is 9.68~J for Nightwatch, 10.23~J for Playwright, 7.49~J for Puppeteer, and 6.51~J for Selenium. Thus the overhead of Selenium with regards to energy consumption is the lowest among the four frameworks.

\subsection{RQ1}

\begin{figure*}[h]
  \centering
  \includegraphics[width=\linewidth]{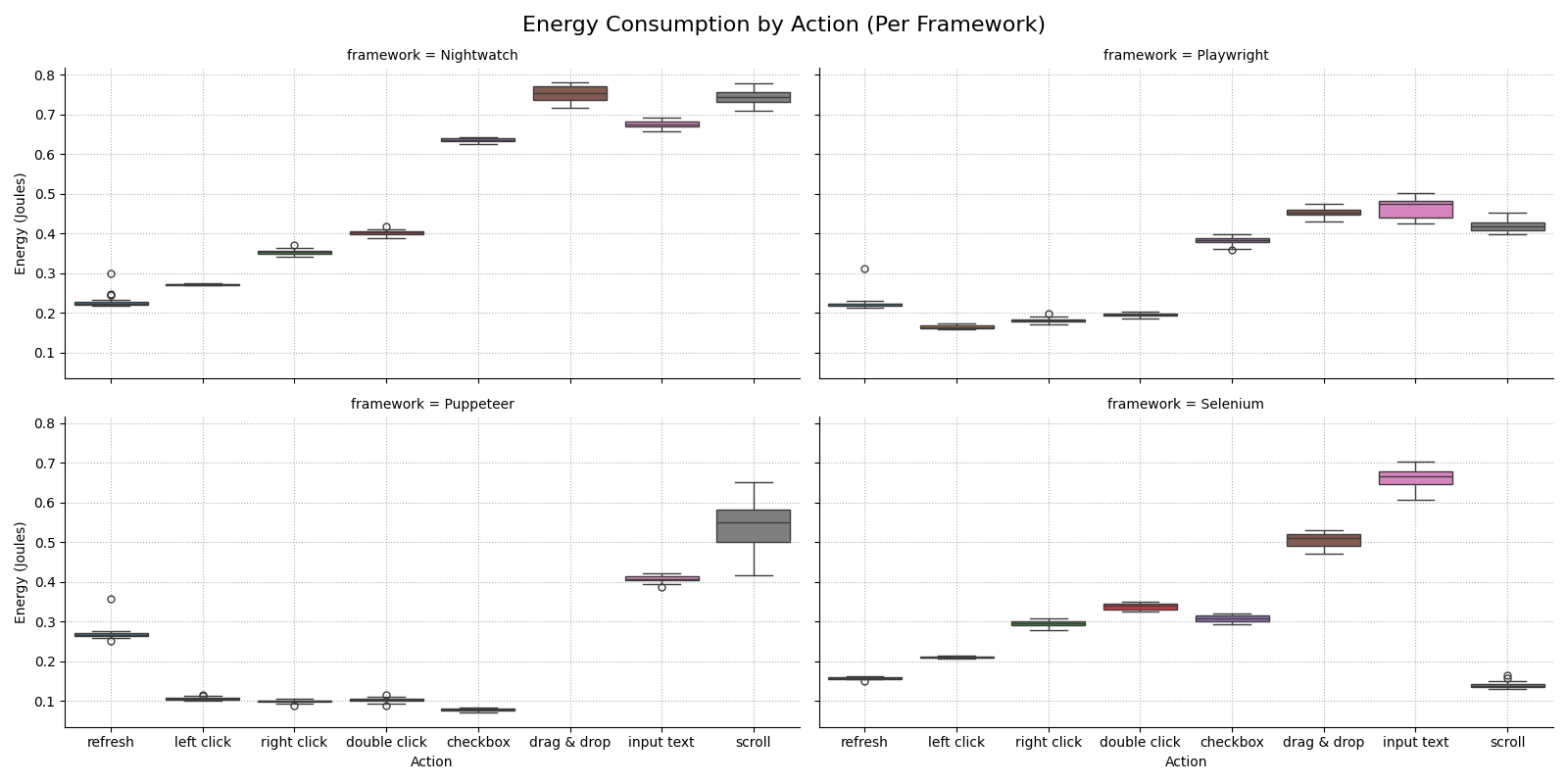}
  \caption{Estimated energy cost per UI action grouped by the different frameworks, where The x-axis lists the different UI actions and the y-axis shows the energy consumption per interaction in joules (J).}
  \label{fig:rq1-boxplot_by_frame}
  \Description{}
\end{figure*}

Figure~\ref{fig:rq1-boxplot_by_frame} shows box plots of energy consumption for the eight UI actions: refresh, left-click, right-click, double-click, checkbox, drag\&drop, input-text, and scroll, across the four frameworks. Within each framework, there is substantial variation in energy usage across actions with some shared tendencies.

\begin{itemize}
    \item{Nightwatch: checkbox, drag\&drop, input-text, and scroll are the most energy intensive actions, while refresh and the clicks are the least energy intensive actions.}
    \item{Playwright: the same pattern is observed, but refresh is more energy intensive than the click actions.}
    \item{Puppeteer: differs in that scroll and input-text are the most costly followed by refresh, while the click actions and checkbox are energy effective and the click actions similar to each other.}
    \item{Selenium: also has input-text and drag\&drop as the most energy intensive actions, while scroll and refresh are energy efficient comparatively.}
\end{itemize}


These findings confirm that UI action type significantly influences energy consumption, even within the same framework. 
To formally assess these differences, we conducted a Kruskal-Wallis test, a non-parametric method suited for comparing multiple independent groups when normality assumptions are violated. The test revealed a statistically significant difference in energy consumption across actions (H = 552.4, p $< 4.2 * 10^{-115}$), indicating that at least one action group differs meaningfully from the others.

\begin{figure}[h]
  \centering
  \includegraphics[width=\linewidth]{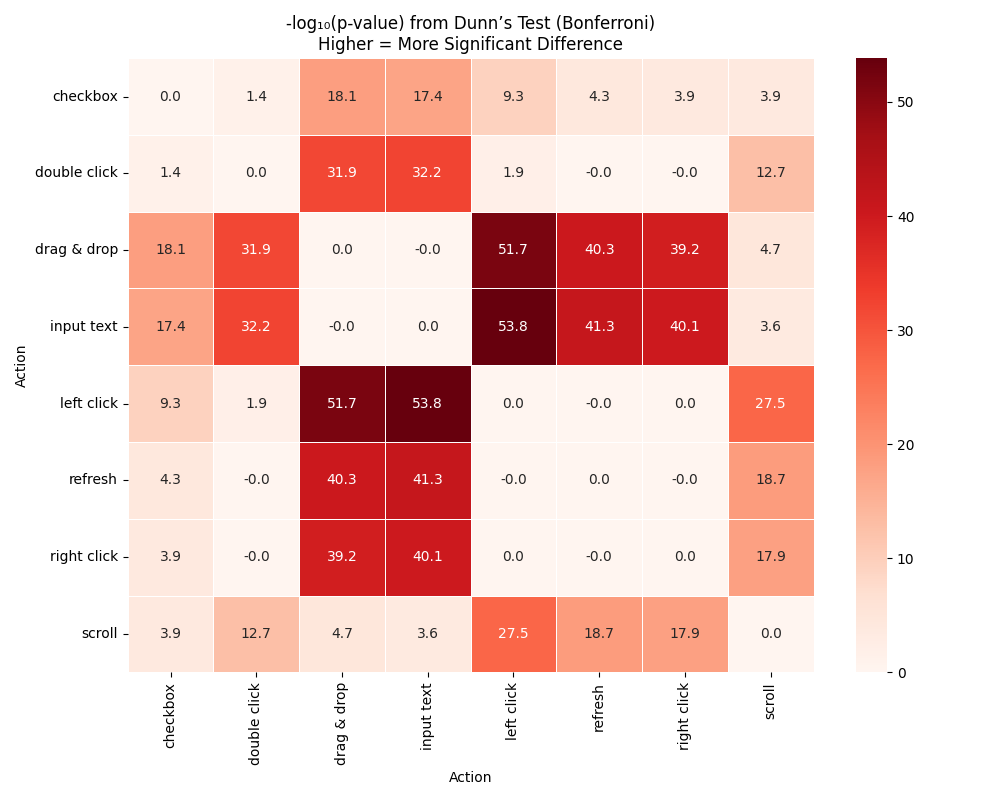}
  \caption{Pairwise significance of energy consumption differences between UI actions based on Dunn’s test with Bonferroni correction. Higher values indicate stronger statistical significance after correction. Darker cells highlight pairs with the largest differences.}
  \label{fig:rq1-dunn-bonferroni}
  \Description{}
\end{figure}

Dunn's test with Bonferroni correction was afterwards done and the results can be seen in figure \ref{fig:rq1-dunn-bonferroni}. Here, it is evident that there are substantial pairwise differences in the energy consumption between the UI actions. The biggest difference here is between left-click and input-text, which has a $-log_{10}(p)$ value of 53.8. The test also indicates that the UI actions can be grouped by their similarity in their energy consumption.

\subsection{RQ2}

\begin{figure*}[h]
  \centering
  \includegraphics[width=\linewidth]{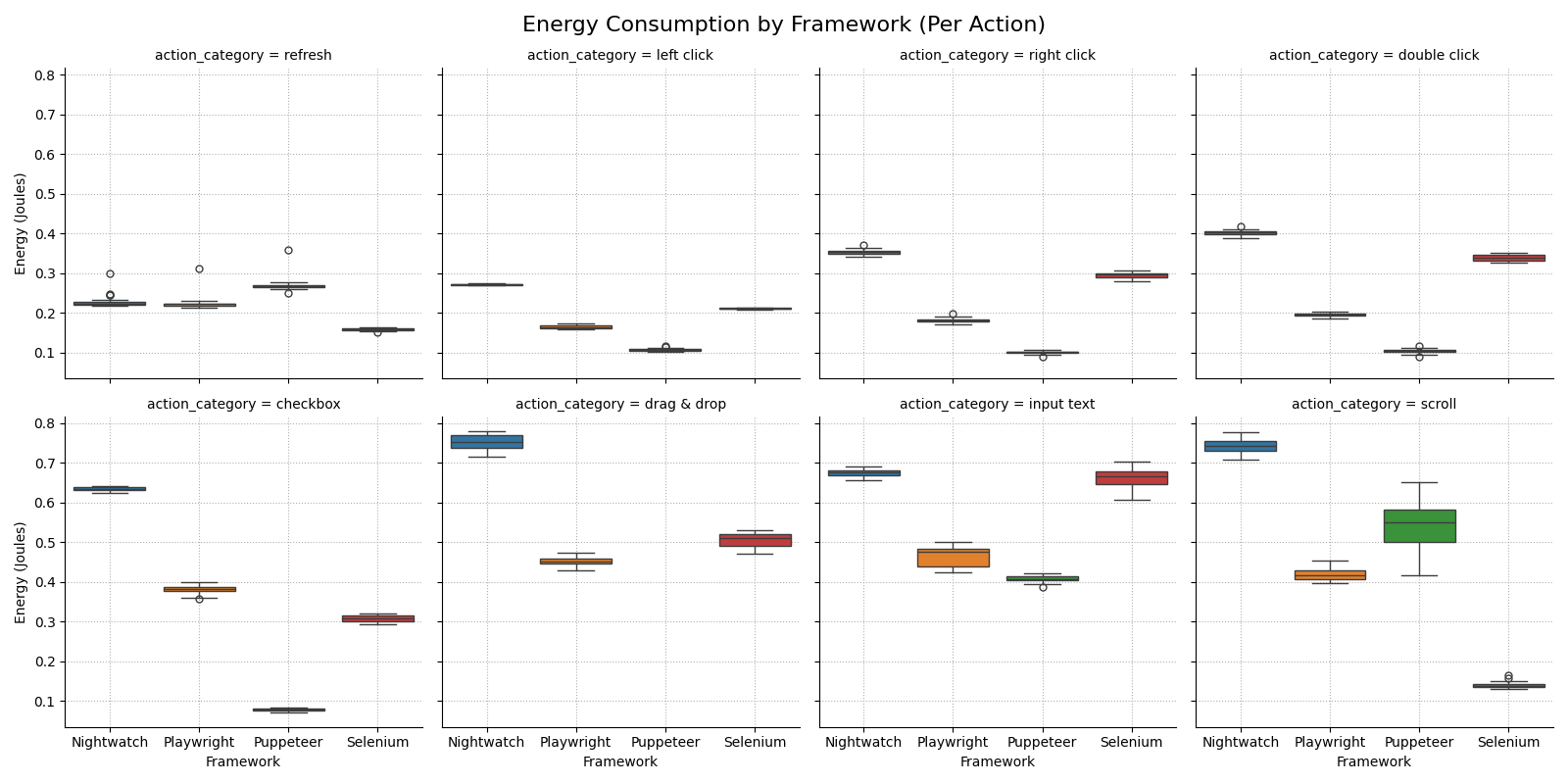}
  \caption{Estimated energy cost per UI action, with the different frameworks represented by different colors. The x-axis lists the different UI automation frameworks and the y-axis shows the energy consumption per interaction in Joules.}
  \label{fig:rq2-boxplot_by_actions}
  \Description{}
\end{figure*}

Figure~\ref{fig:rq2-boxplot_by_actions} compares the energy consumption across frameworks for each UI action. 
The results show that the energy usage for the same action is clearly distinguishable across frameworks. 
The energy results are reported as pairs $(\bar{x}, s_x)$ representing mean and standard deviation respectively in millijoules (mJ).

\begin{description}

  \item[refresh:] Selenium (158 mJ, 3 mJ) is the most energy efficient, whereas Puppeteer (269 mJ, 16 mJ) is the least. Nightwatch (228 mJ, 15 mJ) and Playwright (224 mJ, 16 mJ) perform similarly.

    \item[left-click:] Puppeteer (106 mJ, 3 mJ) is the most energy efficient, followed by Playwright (165 mJ, 4 mJ), Selenium (211 mJ, 1 mJ), and Nightwatch (271 mJ, 1 mJ) as the least efficient.

    \item[right-click:] The ordering follows that of left-click. Interestingly, Puppeteer (100 mJ, 4 mJ) shows almost the same energy use as before, while Playwright (182 mJ, 5 mJ), Selenium (296 mJ, 7 mJ), and Nightwatch (354 mJ, 7 mJ) consume more energy overall.

    \item[double-click:] The same tendency observed from left-click to right-click continues, with Puppeteer (103 mJ, 5 mJ) remaining stable, while a further increase in energy consumption is seen for Playwright (196 mJ, 4 mJ), Selenium (339 mJ, 7 mJ), and Nightwatch (401 mJ, 6 mJ).
    
    \item[checkbox:] Puppeteer (79 mJ, 4 mJ) is again the most energy efficient framework, followed by Selenium (308 mJ, 8 mJ), Playwright (382 mJ, 10 mJ), and Nightwatch (635 mJ, 5 mJ) as the least efficient.

    \item[drag\&drop:] Playwright (453 mJ, 10 mJ) is the most energy efficient framework, closely followed by Selenium (506 mJ, 18 mJ), with Nightwatch (753 mJ, 20 mJ) being the least efficient. Puppeteer does not provide a native drag\&drop action by UI element and is therefore omitted.

    \item[input-text:] Puppeteer (409 mJ, 7 mJ) is the most energy efficient framework, followed by Playwright (464 mJ, 23 mJ), Selenium (662 mJ, 25 mJ), and Nightwatch (675 mJ, 9 mJ) as the least efficient.


    \item[scroll:] Selenium (140 mJ, 7 mJ) shows the best energy efficiency, followed by Playwright (418 mJ, 14 mJ).  Puppeteer (543 mJ, 55 mJ) exhibits the largest variation in energy consumption among all action–framework combinations, while Nightwatch (743 mJ, 17 mJ) has the highest overall energy usage.

\end{description}

\begin{figure}[h]
  \centering
  \includegraphics[width=\linewidth]{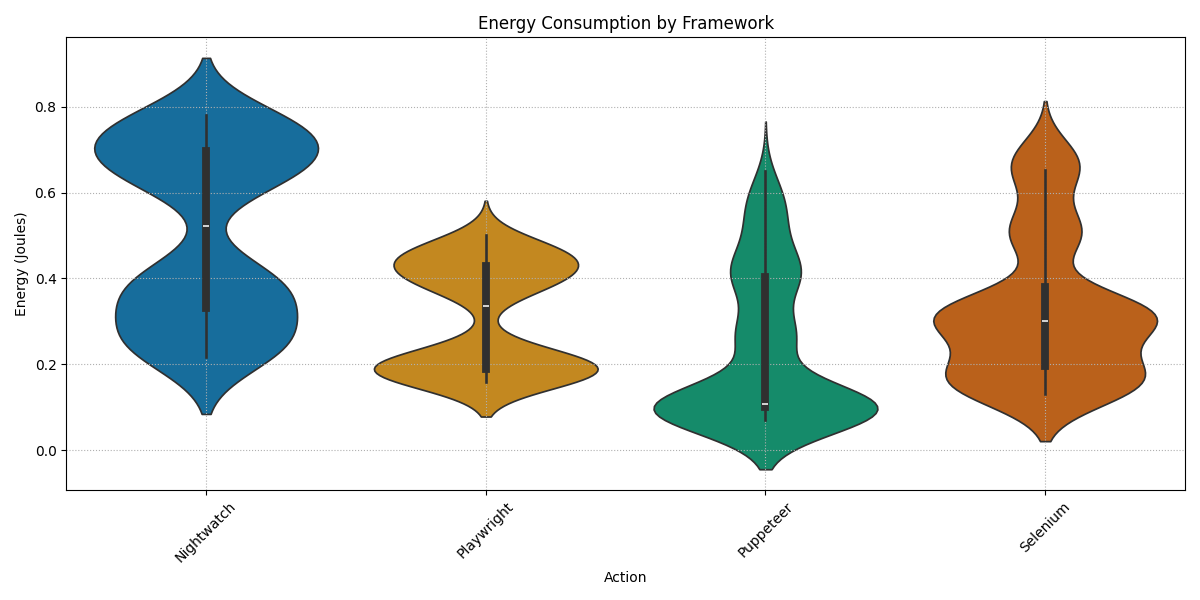}
  \caption{Cumulative energy consumption per framework represented in a violin plot. The x-axis lists the different frameworks and the y-axis shows the energy in Joules per UI action.}
  \label{fig:rq2-violin}
  \Description{}
\end{figure}

Overall, the energy consumption difference between the most and least efficient frameworks varies from approximately twofold to eightfold across actions, with an average reduction potential of about sixfold when selecting the most energy efficient framework.

Figure~\ref{fig:rq2-violin} further supports these findings by showing violin plots of the overall energy consumption per framework. Selenium and Nightwatch exhibit higher medians and wider spreads in energy usage. Puppeteer is generally the most energy efficient framework, although it includes a few less efficient actions. Selenium and Playwright show comparable mean values; however, Selenium displays a more compact distribution toward the lower end of energy consumption. 
Overall, the results indicate that the choice of framework has a measurable impact on the energy consumption of identical UI actions.  Puppeteer and Selenium both show a high density toward lower energy usage, 
whereas Nightwatch and Playwright each exhibit two high-density regions around their mean values.

Kruskal-Wallis test were then done and the results is presented in table \ref{tab:rq2}. For all the actions the H-statistic were high, ranging from (92-130) suggesting substantial differences between the frameworks for all the actions. The very small p-values ($8.72*10^{-21}$ to $6.71*10^{-28}$ means that we can reject the null hypothesis for every action. Which means that the framework in use has a significant effect on the energy consumption for every type of UI action.

\begin{table}[]
\begin{tabular}{lllr}
\hline
action       & H-statistic & p-value         & \multicolumn{1}{l}{Significant? (p \textless 0.05)} \\ \hline
checkbox     & 130.32      & $4.62*10^{-28}$ & True                                                \\
double-click & 130.32      & $4.62*10^{-28}$ & True                                                \\
drag\&drop   & 92.38       & $8.72*10^{-21}$ & True                                                \\
input-text   & 118.61      & $1.53*10^{-25}$ & True                                                \\
left-click   & 130.32      & $4.62*10^{-28}$ & True                                                \\
refresh      & 114.24      & $1.34*10^{-24}$ & True                                                \\
right-click  & 130.32      & $4.62*10^{-28}$ & True                                                \\
scroll       & 129.56      & $6.71*10^{-28}$ & True                                                \\ \hline
\end{tabular}
\caption{Results of Kruskal-Wallis for RQ2}
\label{tab:rq2}
\end{table}

\subsection{RQ3}

To evaluate whether the impact of framework on energy consumption depends on action type, an Aligned Rank Transform (ART) ANOVA was applied, appropriate for interaction testing under non-normal data conditions.
The analysis showed statistically significant main effects for both framework and action:
\begin{description}
    \item[Framework:]  $F(3, 952) = 2151.8, p < 2.22 * 10^{-16}$
    \item[Action:] $F(6, 952) = 5496.4, p < 2.22 * 10^{-16}$
    \item[Interaction:] $F(18, 952) = 2170.9, p < 2.22 * 10^{-16}$
\end{description}
Critically, the interaction effect between framework and action was also highly significant.
These findings confirm that the energy impact of a framework is not consistent across actions, supporting RQ3 and aligning with our earlier visual observations in figure~\ref{fig:rq2-boxplot_by_actions} (e.g.\ non-parallel behavior in interaction plots).

\section{Discussion}
The results show that individual UI actions differ in their energy consumption (\textbf{RQ1}), that the choice of automation framework has a significant impact even for similar actions (\textbf{RQ2}), and that the interaction between framework and action type further amplifies these differences (\textbf{RQ3}). In this section, we interpret these outcomes in light of framework architectures, discuss their methodological and practical implications, and outline directions for future research.

\subsection*{RQ1 Action-Level Energy Differences}

The results for \textbf{RQ1} show clear energy variation between UI actions within each framework.
As observed in our measurements (Section~\ref{results}), the results for RQ1 show distinct energy patterns across the tested UI actions.
In our data, click and right-click consistently exhibited the lowest energy use across frameworks, whereas input-text and scroll required considerably more energy, and refresh dominated overall consumption. These three categories suggests three functional groups that differ in how they engage the browser.

Based on our results, the tested actions naturally cluster into three functional groups.
\begin{description}
\item[Discrete event-driven actions] that trigger short, bounded DOM events and consistently show the lowest energy use. In our case it includes left-click, right-click double-click, and checkbox.
\item[Continuous interactions] that sustain browser activity and DOM updates over time, resulting in higher consumption and variance. In our case, it includes scroll, input-text, and drag\&drop.
\item[System-level actions] initiate full-page reloads with intermediate energy levels, influenced by caching and rendering factors. In our case, it is refresh.
\end{description}
There are, however, two actions with  deviations as follows.
Within the discrete group, checkbox aligns semantically but shows greater variation than other actions, with notably higher energy use in Nightwatch and Playwright.  Since checkbox's energy use aligns to different groups between frameworks, it's placement in the discrete group is debatable: for Selenium and Puppeteer it aligns more closely with the Discrete actions and for Nightwatch and Playwright it aligns better with the energy use in continuous interactions than the discrete actions.

Within the continuous group, scroll follows the same pattern of higher energy consumption, except in Selenium, where it is unexpectedly efficient –it is the least energy consuming action in Selenium.

In general, these pattern aligns well with findings from previous studies.
Cruz et al.~\cite{Cruz2019} reported that continuous user interactions, such as scrolling and typing, maintained sustained CPU and rendering activity, leading to higher energy consumption than discrete events like taps or clicks.

Together, these studies support our observation that energy consumption scales with the duration and complexity of browser activity, forming a consistent behavioral hierarchy across both frameworks and interaction types.

Thus, this grouping emerged empirically from our data and may provide a conceptual lens for interpreting the observed energy differences across frameworks in future research.

\subsection*{RQ2 Framework-Dependent Energy Behavior}
\label{subsec:framework-dependent-energy-behavior}
%
%



Our results confirm that framework choice significantly affects the energy consumed during automated browser interactions (\textbf{RQ2}).  
Although all frameworks execute identical user actions, their internal architectures differ in how commands are dispatched, synchronized, and acknowledged by the browser — differences that manifest as measurable energy variation.


\textbf{Puppeteer}, developed by Google, communicates directly with Chromium-based browsers via the Chrome DevTools Protocol (CDP), bypassing the intermediate driver layer used in WebDriver-based frameworks.
In our results, Puppeteer consistently showed the lowest energy consumption across nearly all actions, indicating that it executes browser commands with less overhead and shorter active durations.
This outcome aligns with expectations based on its architecture: as described in the official documentation~\cite{PuppeteerDocs} and García et al.~\cite{Garcia2024}, Puppeteer’s low-latency WebSocket communication and event-driven control avoid the repeated HTTP request–response cycles characteristic of Selenium and Nightwatch.
Although all frameworks were executed in headless mode in this study, Puppeteer’s streamlined design appears to translate directly into measurable energy savings.

\textbf{Playwright}, developed by Microsoft, extends the DevTools-based control model to support multiple browsers~\cite{PlaywrightDocs}.
It maintains a persistent socket connection to the browser but adds an orchestration layer that unifies browser-specific protocols and ensures deterministic execution across engines~\cite{Garcia2024}.
In our results, Playwright consumed more energy than Puppeteer across most actions but generally less than Selenium and Nightwatch, placing it in the middle range of efficiency.
This pattern is consistent with its design: while the orchestration layer improves reproducibility and cross-browser support, it may also contribute to additional synchronization and CPU activity, which could explain its moderately higher energy use compared with the more direct Puppeteer implementation.

\textbf{Selenium} is the most established automation framework and implements the W3C WebDriver specification~\cite{WebDriverSpec}.
In this framework, client commands are serialized into JSON messages, transmitted over HTTP to a standalone driver process, and then forwarded to the browser~\cite{SeleniumDocs}.
In our results, Selenium consistently consumed more energy than Puppeteer and often more than Playwright. With values in between Playwright and Nightwatch.
This higher energy use coincided with longer execution times and greater variability across action types, particularly for synchronization-intensive interactions.
These patterns are consistent with Selenium’s multi-process, request–response architecture, where blocking waits and driver communication overhead may prolong browser activity and increase overall energy consumption.

\textbf{Nightwatch} builds on Selenium’s WebDriver architecture, adding a JavaScript runtime layer for assertions, logging, and command preprocessing~\cite{NightwatchDocs}. In our results, it exhibited the highest or near-highest energy consumption among the tested frameworks, often exceeding Selenium in both total energy and execution time. Compared to Selenium the longer execution time suggests extra processing or imposed delays before browser execution. This aligns with its design, where the additional abstraction layer offers developer-friendly features but introduces extra processing per command and inherits WebDriver’s communication latency. While this explanation remains tentative, the observed energy pattern is consistent with the expectation that Nightwatch’s added orchestration incurs measurable overhead.

Together, these differences show that energy consumption is shaped not only by the actions under test but also by how frameworks handle synchronization and communication with the browser.
Frameworks optimized for direct, event-driven control (e.g., Puppeteer) tend to exhibit lower energy costs, whereas those emphasizing cross-browser compatibility or higher-level abstractions (e.g., Playwright, Selenium, Nightwatch) incur additional overhead.
These architectural distinctions interact with action complexity, as examined in the following section.

\subsection*{RQ3 Interaction Effects and Interpretation}

The significant interaction between framework and action type indicates that the impact of framework choice is not constant across user interactions.
Actions with complex synchronization requirements (e.g., input-text or drag\&drop) exhibit greater variation between frameworks than simpler events such as clicks.
This pattern suggests that internal design features—such as command queuing, event dispatch latency, and DOM polling frequency—modulate how each framework translates user-level actions into browser operations~\cite{Garcia2024}.

From a measurement perspective, this finding implies that framework  induced bias is \emph{action-dependent}.
This happens because different frameworks handle actions in different ways.
For simple actions such as clicks, all frameworks perform nearly the same sequence of steps.
For more complex actions, such as typing text or scrolling, some frameworks spend extra time waiting, checking, or repeating commands.
These differences affect how long the browser and CPU remain active, which in turn changes the total energy consumed.

Beyond methodological implications, this finding shows that framework overhead and action cost are not independent.
The energy used by a given action depends on how the framework executes it—through its scheduling, waiting, and synchronization behavior.
Therefore, energy measurements should be calibrated for each framework before being used for comparative evaluation.
In other words, aggregated metrics such as total energy per test suite can obscure framework-specific behavior, and per-action energy reporting provides a more reliable basis for reproducible comparisons.

\subsection{Methodological and Measurement Implications}
Although Macedo et al.~\cite{Macedo2020} used identical Selenium scripts to compare the energy efficiency of Chrome and Firefox, our findings indicate that the energy cost of Selenium actions is not constant but depends on how the underlying browser driver implements and synchronizes those actions. Consequently, part of the differences they attributed to browser efficiency may instead originate from browser-specific WebDriver interactions or synchronization delays inherent in Selenium’s execution model.

Similarly, tools such as GreenFrame~\cite{GreenFrame2025} rely on browser automation (currently Selenium) to execute predefined user scenarios and visualize energy consumption at designated milestones. These milestones are designed to help developers identify energy-intensive phases of user flows and guide optimization efforts. However, our results show that the automation framework itself introduces a non-negligible, action-dependent energy overhead. As a result, milestone-based diagnostics may misattribute energy spikes to application code when they actually stem from the framework’s orchestration, or conversely, obscure inefficient interactions when the framework executes them more efficiently. 

This framework dependency highlights a methodological challenge for reproducible energy studies: conclusions drawn from browser-based automation must account for the framework’s own energy profile to ensure valid cross-framework or cross-application comparisons. This observation aligns with the broader concerns raised by Verdecchia et al.~\cite{Verdecchia2025} regarding the need for standardized and reproducible methodologies in software energy research.



\subsection{Internal Validity}
\label{subsec:internal-validity}
\textbf{Power Measurement and Process Isolation}: One limitation concerns the fidelity of power measurements. Although efforts were made to isolate energy usage related to test execution, it remains a challenge to fully eliminate background processes and non-relevant system activity. As a result, there is potential for measurement noise, which may have affected the recorded energy consumption data.

\textbf{Consistency of Action Implementations:}  
Most UI automation frameworks offer multiple syntactic or procedural variants for common actions (e.g., click, scroll, or input-text).
To ensure comparability, we used actions available across all frameworks and implemented them via their standard, developer-facing methods, minimizing the risk that observed differences arise from API usage rather than architectural behavior.

\textbf{Thermal and Temporal Effects:}  
Although system temperature and CPU frequency were monitored, thermal throttling may still have influenced power readings. Because experiments were executed sequentially, earlier runs could experience different conditions than later ones. Randomizing execution order or applying thermal stabilization could help reduce this bias in future work.

\textbf{Repetition and Statistical Confidence:}  
Each measurement was averaged over multiple runs to reduce random variation, with shorter actions repeated more often to counter OS inconsistencies and sensor noise.
Although this improved stability, short actions with low power draw remain sensitive to measurement precision; future work could strengthen robustness through confidence intervals or bootstrapped variance estimation.

\subsection{External Validity}
\label{subsec:external-validity}
While this study provides valuable insights into the energy consumption of UI testing frameworks, several limitations should be acknowledged that may affect the generalisability and interpretation of the results.

\textbf{Hardware Representativeness}: The experiments used Raspberry Pi devices as both client and server, providing controlled and repeatable measurements but limiting generalisability.
Energy patterns on Raspberry Pi may not reflect those of typical end-user hardware (e.g., laptops, desktops, or mobile devices), so future studies should replicate the experiment on more diverse platforms.

\textbf{Headless vs. Headed Execution}: Following standard automated testing procedures, all experiments were conducted in headless mode.
However, it remains uncertain whether this setup accurately reflects the behavior and energy costs of headed (GUI-visible) execution typical in real-world use. Future work should evaluate whether headless measurements reliably approximate energy use in a fully interactive environment

\textbf{Network Context}: All experiments ran on a local network, eliminating variability from internet routing and latency but reducing ecological validity.
Frameworks may behave differently in performance and energy use when interacting with live websites, particularly during page-loading actions affected by real-world network conditions.

\textbf{Framework Coverage}: Although the study initially targeted five frameworks, Cypress was excluded due to execution anomalies, leaving four for analysis and limiting comparison breadth.
Future work could expand coverage by including additional UI automation frameworks.

\textbf{Browser Dependency}: All experiments used the Chromium browser, which is widely supported and representative of many real-world environments.
However, UI action energy performance may differ across browsers, and it remains unclear how well these results generalise to Chrome, Firefox, or Safari.
Future work should compare energy efficiency across browsers using the same action set.

\textbf{Caching and Action Overhead Assumptions}: The use of the refresh action as a baseline raises potential concerns.
If caching lowers energy consumption after the first visit, the assumption that refresh captures full page-load overhead may be invalid.
This issue is particularly important when modeling startup energy costs preceding other UI actions, highlighting the need for a more refined treatment of caching effects.

\subsection{Implications and Future Work}
Our results show that automation frameworks can differ in energy use by more than a factor of six. For a web developer who is interested in testing checkbox funtionality of their website, choosing Puppeteer over Nightwatch would lower their energy overhead from the framework by $635mJ - 79mJ = 556mJ$ for each checkbox action.
This highlights that sustainability in testing depends not only on the web application but also on the tools executing it.
Choosing lighter-weight frameworks in continuous integration pipelines can, therefore, lower testing energy costs and support more sustainable development.

From a methodological perspective, the results show that framework overhead is action-dependent, meaning that each framework introduces its own characteristic energy profile. Energy studies using browser automation should therefore include per-framework calibration or rely on standardized baselines to ensure reproducible and unbiased comparisons across tools.

Looking ahead, the most important next steps are to extend the analysis from isolated UI actions to realistic user flows, capturing cumulative and positional effects, and to refine measurement techniques that better separate framework overhead from the energy consumed by the actions themselves.
Addressing these aspects will enhance the precision and reproducibility of energy measurements in automated web testing and strengthen the methodological basis of energy-aware software engineering.

\section{Conclusion}

The findings from this study have important implications for both researchers and practitioners concerned with energy efficiency in UI automation and web application design.
\textbf{RQ1}: \textit{Do different UI actions vary in their energy consumption within a single UI automation testing framework?} Our data confirms that the energy consumption varies significantly across different UI actions. This implies that energy usage is not uniform across user interactions, and certain actions are inherently more energy intensive. If we assume this pattern extends to end user devices, it suggests that the design of user interaction flows should be approached with energy implications in mind. By minimizing or optimizing energy hungry actions, developers may be able to reduce the overall energy footprint of applications.
\textbf{RQ2} and \textbf{RQ3} focused on the impact of framework selection. \textbf{RQ2}: \textit{Does the choice of UI automation testing framework affect energy use for the same UI action?} The data revealed that the choice of framework significantly affects energy consumption, although the magnitude of the effect varies depending on the specific UI action. This indicates that an energy-aware developer should take energy efficiency into consideration during framework selection.
\textbf{RQ3}: \textit{Is there an interaction between UI automation testing framework and UI action type in terms of energy usage?} The results confirmed a significant interaction effect between the framework and action type. Therefore, the energy impact of a given framework cannot be assessed in isolation as it varies according to the action being executed. In other words, energy consumption is not a simple additive property. Actions behave differently across frameworks, leading to non-uniform energy profiles. Therefore, aggregated energy estimates for user flows composed of multiple actions cannot be assumed to remain constant across frameworks, even if the logical test flow is identical. This has major implications for cross-framework comparisons and portability of test design.

\begin{acks}
This work was supported by the Independent Research Fund Denmark Project no. 2102-00281B. 
We are grateful to Anders Kirkeby for advising during the project, validating the JavaScript implementations, and providing us with the local website.
The authors acknowledge the use of generative AI for assistance on restructuring and shortening of text, and to help identifying related work of relevancy.
\end{acks}

\printbibliography

\end{document}